\documentclass[10pt,twocolumn,letterpaper]{article}

\usepackage{iccv}
\usepackage{times}
\usepackage{epsfig}
\usepackage{graphicx}
\usepackage{amsmath}
\usepackage{amssymb}

\usepackage{enumitem}
\usepackage{subcaption}
\usepackage{caption}
\usepackage{amsfonts}
\usepackage{balance}
\usepackage[acronym]{glossaries}
\usepackage{xspace}
\usepackage{xcolor}
\usepackage{booktabs}
\usepackage{import}
\usepackage{footnote}
\usepackage{dcolumn}
\newcommand{\githublink}{https://github.com/junyanz/pytorch-CycleGAN-and-pix2pix/}

\newacronym{GAN}{GAN}{generative adversarial network}
\newacronym{DCT}{DCT}{discrete cosine transform}
\newacronym{fid}{FID}{Fréchet inception distance}
\newacronym{ssim}{SSIM}{structural similarity index measure}
\newacronym{fm}{FM}{fluorescence microscopy}

\usepackage[breaklinks=true,bookmarks=false]{hyperref}

\iccvfinalcopy 

\def\iccvPaperID{26} 

\newcolumntype{P}{D{p}{\, \pm \,}{4,2}}

\begin{document}

\title{Focus on Content not Noise: Improving Image Generation for Nuclei Segmentation by Suppressing Steganography in CycleGAN}

\author{%
\begin{tabular}{c}
    Jonas Utz$^{1}$ \quad Tobias Weise$^{1,2}$ \quad Maja Schlereth$^{1}$ \quad Fabian Wagner$^{2}$ \\ Mareike Thies$^{2}$ \quad Mingxuan Gu$^{2}$ \quad Stefan Uderhardt$^{3,4,5}$ \quad Katharina Breininger$^{1}$ \\
      \small $^{1}$Department Artificial Intelligence in Biomedical Engineering, Friedrich-Alexander-Universität Erlangen-Nürnberg, Germany \\ 
      \small $^{2}$ Pattern Recognition Lab, FAU Erlangen-Nürnberg, Germany \\
      \small $^{3}$ Department of Medicine 3 - Rheumatology and Immunology, FAU Erlangen-Nürnberg, Germany \\
      \small $^{4}$ Deutsches Zentrum für Immuntherapie, FAU Erlangen-Nürnberg, Germany \\
        \small $^{5}$ Exploratory Research Unit, Optical Imaging Centre Erlangen, FAU Erlangen-Nürnberg, Germany \\
    \tt\small jonas.utz@fau.de
\end{tabular}
}

\maketitle

\begin{abstract}
Annotating nuclei in microscopy images for the training of neural networks is a laborious task that requires expert knowledge and suffers from inter- and intra-rater variability, especially in fluorescence microscopy. Generative networks such as CycleGAN can inverse the process and generate synthetic microscopy images for a given mask, thereby building a synthetic dataset. However, past works report content inconsistencies between the mask and generated image, partially due to CycleGAN minimizing its loss by hiding shortcut information for the image reconstruction in high frequencies rather than encoding the desired image content and learning the target task. In this work, we propose to remove the hidden shortcut information, called steganography, from generated images by employing a low pass filtering based on the \gls{DCT}. We show that this increases coherence between generated images and cycled masks and evaluate synthetic datasets on a downstream nuclei segmentation task. Here we achieve an improvement of $5.4$ percentage points in the F1-score compared to a vanilla CycleGAN. Integrating advanced regularization techniques into the CycleGAN architecture may help mitigate steganography-related issues and produce more accurate synthetic datasets for nuclei segmentation.
\end{abstract}
\glsresetall
%
\section{Introduction}\label{sec:intro}
\begin{figure}[b]
    \centering
    \begin{subfigure}[b]{0.22\textwidth}
		\centering
		\includegraphics[width=3.7cm,height=3.7cm]{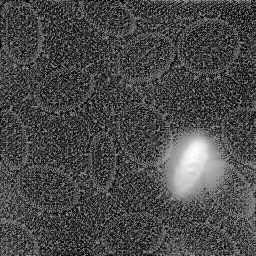}
		\caption{Generated nuclei}
		\label{fig:ahe_fake}
	\end{subfigure}
    \hfill
	\begin{subfigure}[b]{0.22\textwidth}
		\centering
		\includegraphics[width=3.7cm,height=3.7cm]{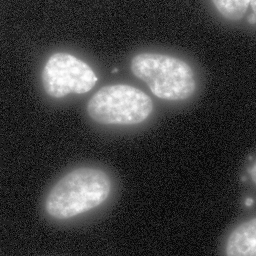}
		\caption{Real nuclei}
		\label{fig:ahe_real}
	\end{subfigure}
    \caption{\ref{fig:ahe_fake} image with steganography amplified by adaptive histogram equalization. The generated image contains information in seemingly empty regions. \ref{fig:ahe_real} Processed real nuclei image for reference}
    \label{fig:stego_ahe}
\end{figure}
\begin{figure}[t]
	\centering
	\includegraphics[width=0.48\textwidth]{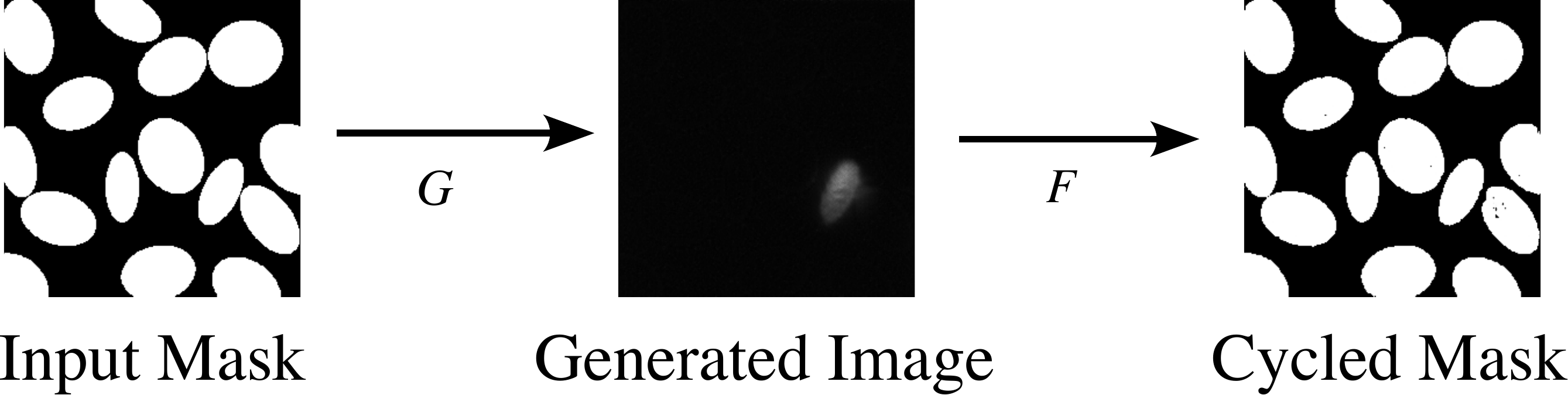}
	\caption{Severe example of CycleGAN steganography: Despite only one nucleus being visible in the generated image, generator $F$ can reconstruct the input mask almost perfectly. This is possible due to the embedding of a compressed version of the input mask in the generated image as a high-frequency low-amplitude signal by generator $G$.}
	\label{fig:stego}
\end{figure}

 
Detecting individual nuclei and cellular structures in microscopy images is a common task in biomedical image analysis and provides insights into cellular processes. Deep neural networks have been successfully used to automate nuclei detection but are typically trained in a supervised manner with a paired dataset of nuclei images and manual labels~\cite{stardist,cellpose}. However, acquiring manual labels is time-consuming and prone to errors and inconsistencies, especially in \gls{fm} images, e.g., due to clustered nuclei, bleed-through, and other artifacts. Unsupervised methods present an alternative, with two approaches being commonly used: a two-stage approach where a generative model is used to generate a synthetic dataset for training a supervised segmentation network and a one-stage approach where a generative model is directly used to perform nuclei segmentation. In both cases, the generative model is trained with an unpaired dataset of real microscopy images and synthetic masks which can be obtained from a suitable synthesis model such as ellipses \cite{boehland}. The model is then trained to perform unpaired image-to-image translation between masks and nuclei images.
A commonly chosen approach is the CycleGAN architecture, introduced by Zhu \etal~\cite{cygan}. A CycleGAN employs two generators, each dedicated to translate between two image domains and two discriminators for the respective domains. A cycle-consistency loss for the generators enforces consistency between an original image and its cycled version to ensure the learned mapping is a bijection. Essentially, the \textit{cycle loss} ensures that the generators retain the image content and only transfer the domain (\eg, from mask to nuclei). However, this constraint does not necessarily lead to a content similarity between the input and the generated image in the other domain. As a shortcut to minimize the cycle-consistency loss, generators can embed a compressed version of the input in the generated image as a high-frequency low-amplitude signal to create a perfect reconstruction of the input image \cite{chu_stego,porav_stego}.
This emerging property has been denoted as CycleGAN \emph{steganography} since the generator actively hides information from the discriminator. This can cause a mismatch between the source image and the generated image leading to label noise in the context of synthetic dataset generation. The prominent effect observed in nuclei images could be attributed to the minimal influence of missing instances on the discriminator's output. Since the discriminator's role is to classify whether an image appears real, the absence of certain nuclei instances does not considerably impact the realism of the microscopy image. An impression of the CycleGAN steganography effect can be obtained from Figure~\ref{fig:stego}. By performing adaptive histogram equalization, which is an image processing technique that enhances contrast by redistributing pixel intensities based on local regions the steganography signal we visualize the steganography signal in Figure~\ref{fig:stego_ahe}. For reference, we apply the adaptive histogram equalization not only to the generated image but also a real nuclei image. The unprocessed version of Figure~\ref{fig:ahe_fake} is also used in Figure~\ref{fig:stego}.
The CycleGAN steganography property was first reported by Chu \etal in \cite{chu_stego}. Since the hidden signal is of low amplitude with a high frequency, it has  characteristics similar to noise. Motivated by this observation, Porav \etal \cite{porav_stego} propose using DnCNN, an end-to-end differentiable image denoising network, between the two generators to remove hidden information. The denoising network is only applied for the cycle-consistency loss and is frozen during CycleGAN training. They evaluate their method on a day-to-night style transfer task with semantic segmentation. L\"ohdefink and Fingscheidt \cite{noise} try to corrupt the hidden signal and break undesired information exchange between the generators by adding Gaussian noise or discretization noise to the generated images. They evaluate their approach on a supervised semantic segmentation of driving scenes. \\
In this paper, we generate a fully synthetic dataset of \gls{fm} images of nuclei with a CycleGAN using transformed ellipses as input masks. To mitigate the before-mentioned CycleGAN steganography problem, we propose to use the \gls{DCT} to perform low-pass filtering of the generated images to remove the hidden signal adaptively.  While we still observe a discrepancy in content between the input mask and generated nuclei images using our approach, the cycled mask matches the content of the generated image. This allows us to use the generated images with their cycled masks as training data for the supervised training of segmentation networks. We evaluate our approach by testing only with synthetic data trained networks on real data. We observe an increased average precision when DCT filtering is used due to the improved alignment between the mask and generated image.
To summarize, the contributions of our work are:
\begin{itemize}
	\item We compare different CycleGAN steganography removal techniques on \gls{fm} images.
	\item We propose an efficient way to remove steganography from CycleGAN images by introducing an additional filter step, based on the \gls{DCT}.
	\item We analyze the impact of CycleGAN steganography on downstream segmentation performance on two public \gls{fm} nuclei datasets and show improved performance when using our proposed method.
\end{itemize}
%
\section{Methods}\label{sec:methods}
\begin{figure*}[t]
    \centering
    \includegraphics[width=\textwidth]{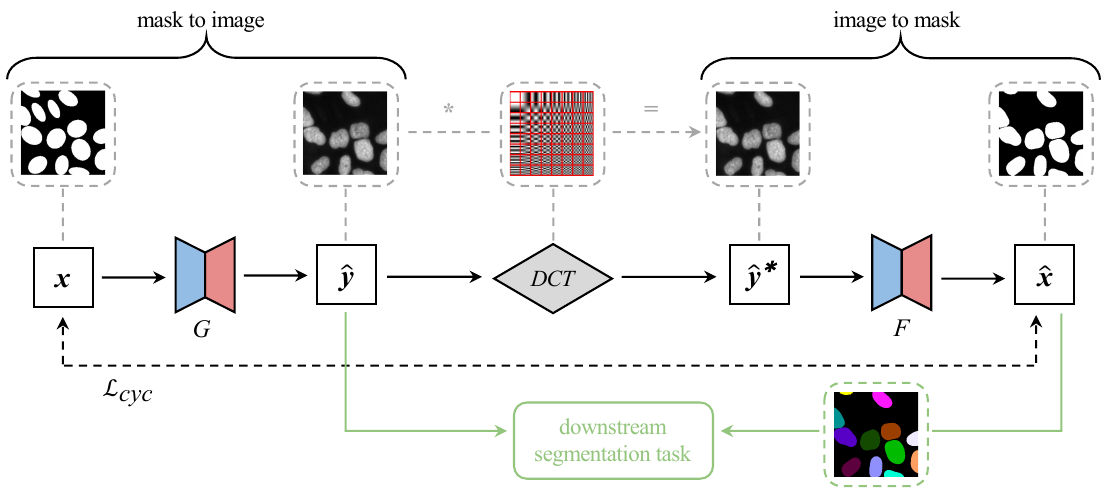}
    \caption{Extended CycleGAN architecture featuring a DCT-based low-pass filter, effectively removing undesired steganography from generated images. The downstream segmentation network evaluates the model's performance and is trained on generated images and cycled masks.}
    \label{fig:schematic}
\end{figure*}
\subsection{CycleGAN}
Our method is based on CycleGAN\cite{cygan} which can learn a translation between two domains, such as the domain of masks $X$ and nuclei images $Y$, with individual samples from the domains denoted as $\boldsymbol{x}$ and $\boldsymbol{y}$ respectively. This translation is performed by two generator networks $G: X \rightarrow Y$ and $F: Y \rightarrow X$. Both generators are trained with an adversarial loss using discriminator networks $D_X$ and $D_Y$ for each domain. For generator $G$, which receives a mask image $\boldsymbol{x}$ and generates a fake nuclei image $\boldsymbol{\hat{y}}$, the adversarial loss $\mathcal{L}_{\mathrm{adv}}$ is defined as:
\begin{equation}
\begin{aligned}
\mathcal{L}_{\mathrm{adv}} & =\mathbb{E}_{\boldsymbol{y} \sim p_{\text {data }}(\boldsymbol{y})}\left[\log D_Y(\boldsymbol{y})\right] \\
& \phantom{={}} +\mathbb{E}_{\boldsymbol{x} \sim p_{\text {data }}(\boldsymbol{x})}\left[\log \left(1-D_Y(G(\boldsymbol{x}))\right]\right.
\end{aligned}
\end{equation} 
The adversarial loss for generator $F$ is computed analogously. An additional \textit{cycle consistency loss} is used to ensure that the mapping is bijective by enforcing $F(G(\boldsymbol{x})) \approx \boldsymbol{x}$ and $G(F(\boldsymbol{y})) \approx \boldsymbol{y}$. In other words, a sample should be close to its original after being consecutively processed by both generators. The cycle-consistency loss for generator $G$ ($F$ again analogously) is formulated as:
\begin{equation}
\begin{aligned}
\mathcal{L}_{\text{cyc}} = \mathbb{E}_{\boldsymbol{y} \sim p_{\text{data }}(\boldsymbol{y})}\left[\|G(F(\boldsymbol{y}))-\boldsymbol{y}\|_{1}\right].
\end{aligned}
\label{eq:cycleloss}
\end{equation}      
\subsection{DCT filtering} 
\begin{figure}
    \centering
	\begin{subfigure}[b]{0.15\textwidth}
		\centering
		\includegraphics[width=2.5cm,height=2.5cm]{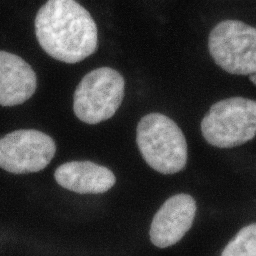}
		\caption{Generated image}
		\label{fig:gen_orig}
	\end{subfigure}
	\begin{subfigure}[b]{0.15\textwidth}
		\centering
		\includegraphics[width=2.5cm,height=2.5cm]{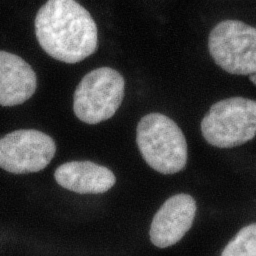}
		\caption{Filtered image}
		\label{fig:gen_filt}
	\end{subfigure}
	\begin{subfigure}[b]{0.15\textwidth}
		\centering
		\includegraphics[width=2.5cm,height=2.5cm]{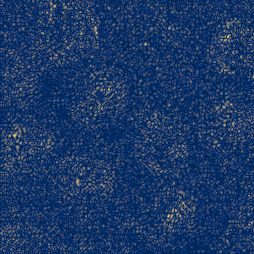}
		\caption{Difference image}
		\label{fig:gen_diff}
	\end{subfigure}
	\caption{Generated images (a) filtered with our proposed DCT filtering are almost indistinguishable since the filter is designed only to remove steganography. In (c) the absolute difference between the filtered and the generated image is shown.}
	\label{fig:filter_comparison}
\end{figure}

The DCT algorithm is best known in the context of the JPEG image compression algorithm \cite{jpeg}, which removes high frequencies from images for compression. We propose eliminating the high-frequency pixel-level patterns \cite{chu_stego} used by CycleGANs to hide  information by setting coefficients associated with higher frequencies to zero.
The \gls{DCT} is a bijective mapping between an input $I$ and the frequency domain. In this work, $I$ refers to the generated image by CycleGAN, \ie, $\boldsymbol{\hat{y}}$ or $\boldsymbol{\hat{x}}$. In the following equations, $I_{xy}$ represents the pixel intensity at position $(x, y)$ in the image. The coefficients $C$ of the DCT are computed on the image $I$ of $N \times M$ pixels as follows:
\begin{equation}
\begin{aligned}
		\operatorname{DCT}(i, j)= 
	\alpha(i) \alpha(j) \sum_{x=0}^{N-1} \sum_{y=0}^{M-1} I_{xy}hv, \\
	\text{where } h = \cos \frac{\pi(2 x+1) i}{2 N}, v = \cos \frac{\pi(2 y+1) j}{2 M}.
\end{aligned}
\end{equation}
The functions $h$ and $v$ weight the contributions of different frequency components in the image $I_{xy}$ to calculate the DCT coefficient at position $(i, j)$.The scaling factor $\alpha$ can be used to make the transform orthonormal:
\begin{equation}
	\alpha(k)= \begin{cases}\frac{1}{\sqrt{2}}, & \text{if } k=0 \\ 1, & \text {otherwise}\end{cases}.
\end{equation}
When computed on $2D$ images of size $N \times M$, $N\times M$ coefficients $C^{M\times N}$ are obtained. In order to eliminate specific frequency bands from the image, we applied a technique of setting corresponding DCT coefficients to zero in the frequency domain. This was achieved by element-wise multiplication of the DCT coefficients matrix with a mask matrix, where the mask contained zeros at the positions corresponding to higher frequencies. The purpose of this step is to remove steganography from the generated images, as the steganography signal, introduced into the images by the generators to ease up reconstruction, is of low amplitude and high frequency. Specifically, we performed low-pass filtering by nullifying the coefficients corresponding to the $50\%$ highest frequencies. To obtain an image $\boldsymbol{\hat{y}^*}$ or $\boldsymbol{\hat{x}^*}$ from the filtered coefficients $C^{*}$, the inverse \gls{DCT} is applied by computing:
\begin{equation}
	\begin{aligned}
		\operatorname{iDCT}(i, j)=
		\sum_{i=0}^{N-1} \sum_{j=0}^{M-1} \alpha(i) \alpha(j) C^{*}(i, j)hv.
	\end{aligned}
\end{equation}
Naturally, the filtering does not only affect the hidden signal but any high-frequency image information, such as foci inside of the nuclei. However, we argue that for reconstructing masks from nuclei images, these high-frequency features are not mandatory, and the differences are subtle. A side-by-side comparison between an original image and a \gls{DCT} filtered image and their absolute difference is depicted in Figure~\ref{fig:gen_diff}. As the cycle loss mainly promotes the CycleGAN steganography, we apply the filtering only to images used for computation of the cycle loss, \ie, the adversarial losses are computed on the unfiltered generator output. Figure~\ref{fig:schematic} gives an overview of the CycleGAN data flow with our proposed DCT filtering for synthetic dataset generation.
\subsection{Mask synthesis}\label{sec:mask_synth}
B\"ohland \etal show that ellipses are a simple but suitable model for representing the shape of nuclei \cite{boehland} and can be used to create synthetic label images. However, careful parameterization of the ellipses is required to match the properties of the nuclei of real microscopy images. A substantial deviation between synthetic and real distribution might result in CycleGAN being unable to learn the mapping correctly. The standard equation for an ellipse is
\begin{equation}
\frac{x^2}{a^2}+\frac{y^2}{b^2}=1,
\end{equation}
with the semi-major axis $a$ and the semi-minor axis $b$.
To express how elongated an ellipse is, we use the eccentricity $e$, which is for $a > b$ defined as:
\begin{equation}
e=\sqrt{1-\left(\frac{b}{a}\right)^2}.
\end{equation}
For mask synthesis, we randomly sample the eccentricity in the range $[0.4, 0.9]$ to compute the minor axis $b$ with:
\begin{equation}
b=\sqrt{1-e^2} a.
\end{equation} 
The size of the semi-major axis $a$ is randomly sampled from a uniform range which is determined on a per-dataset basis. Similarly, the number of nuclei for each image is determined.  A random rotation with a random angle from a uniform distribution in the range [0°, 179°] is applied to each nucleus individually. Nuclei are placed in the synthetic mask images so that they may touch but never overlap.  
\section{Experiments}
\subsection{Evaluation setup}
\begin{figure}[htbp]
	\centering
	\includegraphics[width=0.35\textwidth]{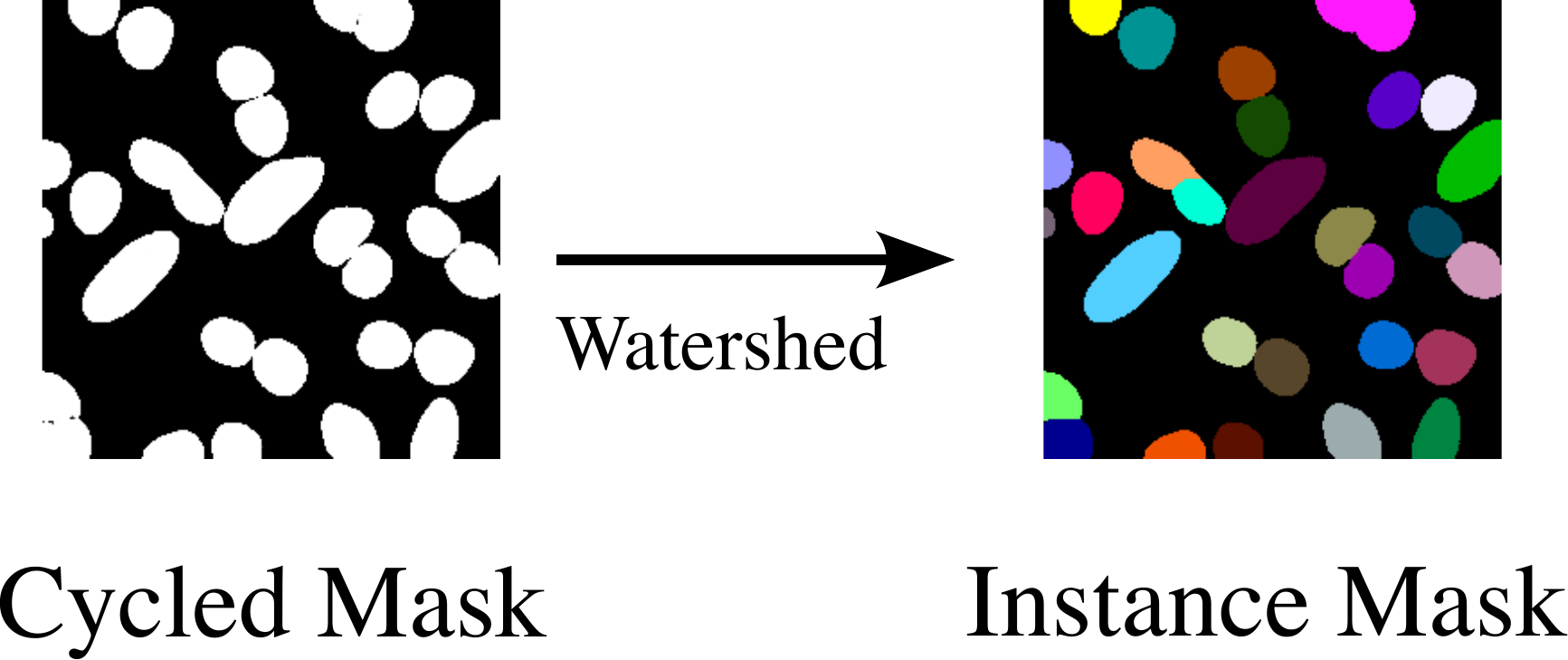}
	\caption{To convert the binary cycled masks into instance masks that can be used to train segmentation networks, we apply the watershed algorithm on the distance-transformed cycled masks.}
	\label{fig:postproc}
\end{figure}

To evaluate the influence of DCT filtering on CycleGAN training, we generate synthetic datasets using a CycleGAN with DCT filtering and an unmodified CycleGAN as a baseline. Additionally, we implement two methods from literature: the denoising approach (\textit{DnCNN}) from \cite{porav_stego} and the noise-injection approach (\textit{Noise}) using Gaussian noise as proposed in \cite{noise}. Instead of using the synthetic input mask from our mask synthesis model and the generated microscopy image by CycleGAN for our synthetic dataset, we use the cycled input mask along with the generated microscopy image. This is motivated by the fact that our method aims to increase correspondence between generated image and cycled mask by filtering the generated image. Additionally, ellipses are a suitable but imperfect approximation of the nuclei shape. If the CycleGAN deviates slightly from this approximation, the cycled mask allows adaptation to these deviations. This is under the assumption that CycleGAN steganography was removed successfully. However, the cycled masks generated by CycleGAN are close to binary masks that might contain minor imperfections. To convert the binary masks into instance masks, we close small holes in the masks and compute the Euclidean distance transform of the binary mask and then apply Watershed segmentation with morphological operators. An example of this post-processing step is visualized in Figure \ref{fig:postproc}. The synthetic dataset is then utilized for training the nuclei segmentation network StarDist \cite{stardist}. Subsequently, the trained segmentation network is applied to real microscopy data for testing. We use StarDist's matching function to compute precision and recall with IoU thresholds of $\tau = 0.5$ and $\tau = 0.75$. Additionally, we report the F1 score. Each dataset is generated three times with random initialization and results are averaged.
\subsection{Datasets}
\begin{figure*}[ht]
    \centering
    \begin{subfigure}{0.22\textwidth}
        \centering
        \includegraphics[width=\textwidth]{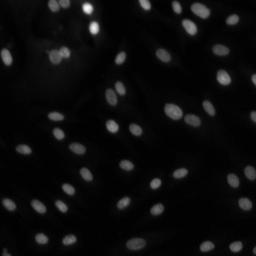}
    \end{subfigure}
    \hfill
    \begin{subfigure}{0.22\textwidth}
        \centering
        \includegraphics[width=\textwidth]{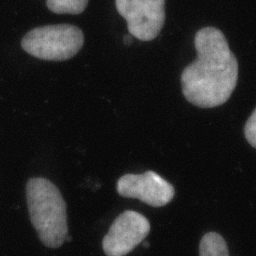}
    \end{subfigure}
    \hfill
    \begin{subfigure}{0.22\textwidth}
        \centering
        \includegraphics[width=\textwidth]{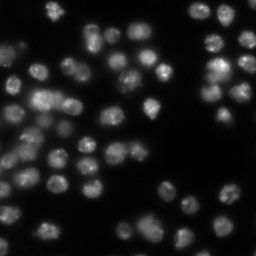}
    \end{subfigure}
    \hfill
    \begin{subfigure}{0.22\textwidth}
        \centering
        \includegraphics[width=\textwidth]{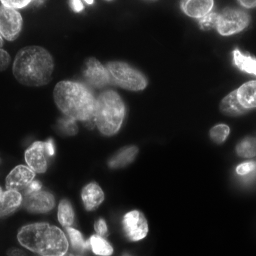}
    \end{subfigure}
    \hfill
    \caption{Sample images from DSB dataset}
    \label{fig:dsb_samples}
    \vspace{0.5cm} 
    \begin{subfigure}{0.22\textwidth}
        \centering
        \includegraphics[width=\textwidth]{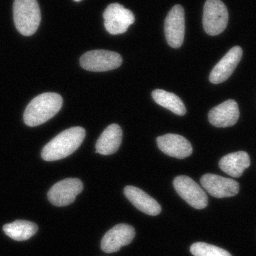}
    \end{subfigure}
    \hfill
    \begin{subfigure}{0.22\textwidth}
        \centering
        \includegraphics[width=\textwidth]{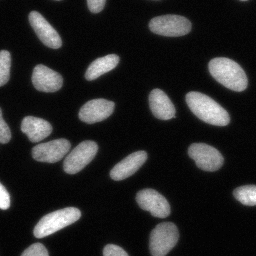}
    \end{subfigure}
    \hfill
    \begin{subfigure}{0.22\textwidth}
        \centering
        \includegraphics[width=\textwidth]{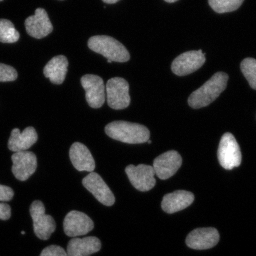}
    \end{subfigure}
    \hfill
    \begin{subfigure}{0.22\textwidth}
        \centering
        \includegraphics[width=\textwidth]{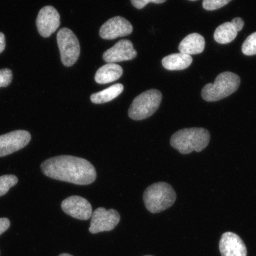}
    \end{subfigure}
    \hfill
    \caption{Samples from BBBC039v1 dataset}
    
    \label{fig:3bc_samples}
\end{figure*}
To evaluate our method, we train CycleGAN on two public \gls{fm} nuclei datasets from the Broad Bioimage Benchmark Image Collection \cite{bbbc}. For each nuclei image, we normalize pixel values into the range $[-1,1]$ and extract random crops of $256 \times 256$ pixels. To create synthetic labels, we use the mask synthesis model described in Section~\ref{sec:mask_synth}. The parameters for mask synthesis can be obtained from Table~\ref{tab:dsb_params} for the DSB dataset and Table~\ref{tab:3bc_params} for dataset BBBC039v1. They were determined based on a small set of three selected nuclei images for each dataset individually. During the training of CycleGAN, new random masks are created for each batch. After the training of CycleGAN, a fixed set of $1500$ synthetic masks is used to generate nuclei images as a training set for the downstream nuclei segmentation network.  
\subsubsection{Data Science Bowl (DSB) 2018 Nuclei Dataset} 
The Data Science Bowl (DSB) 2018 Nuclei Dataset was released for the 2018 Data Science Bowl Kaggle challenge and is listed in the Broad Bioimage Benchmark Collection as BBBC038v1. The dataset contains thousands of nuclei obtained from various organisms, including humans, mice, and flies. Nuclei in the dataset appear in different contexts and states, \eg, cell division, genotoxic stress, and differentiation. While the original dataset contains a variety of treatments (\eg, hematoxylin and eosin stained nuclei), only \gls{fm} images were used in this work. The dataset contains two labeled subsets, \textit{stage1\_train} with 538 \gls{fm} images and \textit{stage1\_test} with 53 \gls{fm} images. Additionally, the dataset contains an unlabelled subset, \textit{stage2\_test} of which we select 667 \gls{fm} images. We use \textit{stage1\_train} and \textit{stage2\_test} for CycleGAN training. The downstream nuclei segmentation networks were tested on \textit{stage1\_test}. Figure~\ref{fig:dsb_samples} visualizes the variety of the dataset using three example images.
\begin{table}
\caption{Parameters of the mask synthesis model of DSB dataset. Each synthetic mask's parameters are chosen from random uniform intervals denoted below. By using multiple subsets of parameter combinations, we allow the mask synthesis model to reflect the diverse characteristics of the DSB dataset.}
\label{tab:dsb_params}
\centering
\begin{tabular}{ccc}
	\hline Length major axis & Nuclei count & Eccentricity $e$\\
	\hline
	$(5,10)$ & $(1,150)$ & $(0.4, 0.9)$ \\
	$(10,15)$ & $(1,40)$ & $(0.4, 0.9)$ \\
	$(15,20)$ & $(1,40)$ & $(0.4, 0.9)$ \\
	$(20,25)$ & $(1,40)$ & $(0.4, 0.9)$ \\
	$(25,30)$ & $(1,20)$ & $(0.4, 0.9)$ \\
	$(30,35)$ & $(1,20)$ & $(0.4, 0.9)$ \\
	\hline
\end{tabular}
\end{table}
\subsubsection{BBBC039v1 dataset} 
The dataset BBBC039v1 is a \gls{fm} dataset, and the images were obtained as part of a high-throughput chemical screen on U-2 OS cells by Caicedo \etal \cite{unetdataset}. The dataset presents a variety of nuclei shapes, different degrees of clustering, and consists of a total of 230000 single nuclei, which were manually annotated. However, compared to the DSB dataset, it exhibits much less variability in shape and nuclei style, as all images were obtained from the same experiment. The dataset holds 200 unique images, each of size $520 \times 696$ pixels. Of these, 150 were used for CycleGAN training. The remaining 50 images were used as the test set for the downstream nuclei segmentation network. Exemplary images are shown in Figure~\ref{fig:3bc_samples}.
\begin{table}[htbp]
\caption{Parameters of the mask synthesis model of BBBC039v1 dataset. Each synthetic mask's parameters are chosen from random uniform intervals denoted below. Compared to DSB, the BBBC039v1 dataset is less diverse concerning nuclei shape and distribution, reflected by our selected parameters.}
\label{tab:3bc_params}
\centering
\begin{tabular}{ccc}
	\hline Length major axis & Nuclei count & Eccentricity $e$\\
	\hline
	$(10,20)$ & $(20,60)$ & $(0.6, 0.9)$ \\
	$(20,40)$ & $(20,30)$ & $(0.6, 0.9)$ \\
	\hline
\end{tabular}
\end{table}
\subsection{Network training}
\subsubsection{CycleGAN}
All compared CycleGAN variants are based on the network architecture described in the original CycleGAN paper \cite{cygan} and we used the official PyTorch implementation.\footnote{Available at \href{\githublink}{\githublink}.} We trained all networks for 100 epochs with a learning rate of 2e-4, which was linearly decayed to 0 for another 100 epochs, using an Adam optimizer ($\beta_1 = 0.5$ and $\beta_2 = 0.999$) with a batch size of $1$. To stabilize the training, the discriminator was updated from an image pool of 50 images \cite{imagepool}. The weight $\lambda$ for the cycle-consistency loss was set to $\lambda_F = 15.0$ for generator $F$ and to $\lambda_G = 10.0$ for generator $G$. We set $\lambda_G$ lower since the mapping from masks to nuclei images is not unique, as it is not directly possible to infer the texture from a nucleus based on a binary mask. For our comparison with \cite{porav_stego}, we use the reference implementation of \textit{DnCNN} \cite{dncnn}, which is trained for blind Gaussian denoising with a large range of noise levels $(\sigma \in[0,55])$. The denoising network weights were frozen during training and inference runs. Denoising is only applied for computation of the cycle loss. 
\subsubsection{StarDist}
We use StarDist \cite{stardist} as downstream nuclei segmentation network. It is a popular choice in the field due to its effectiveness, versatility, and state-of-the-art nuclei segmentation performance. However, in this work, the focus is not on achieving the best possible result but on investigating the applicability of differently generated synthetic datasets for a downstream task. We train StarDist for 50 epochs with a batch size of 16 and a learning rate of $3 \times 10^{-4}$. We increased the number of rays used for the star-convex polygon representation from the default 32 to 64 in order to obtain more detailed polygon shapes. All remaining parameters of StarDist were left unchanged. 
\label{sec:experiments}
\section{Results}
\subsection{Assessment of image fidelity}
While the primary goal of our proposed methods is to improve the downstream task of nuclei segmentation, we report the results of the fidelity comparison for generated and original images in Table~\ref{tab:fidelity} for completeness. To this end, we compute the \gls{fid} as proposed in \cite{cleanfid}. The CycleGAN with DCT filtering yields the best FID score ($96.4$ for both datasets. The low standard deviation ($\pm 0.7$ for the three repetitions indicates that DCT filtering can also help to produce more stable results. We generally observe unusually high FID scores among all datasets. However, the \gls{fid} score computation is based on a feature extractor trained on natural images. This representation might not be optimal for quality assessment of \gls{fm} images, leading to higher scores and limited validity.
\begin{table}[htbp]
\caption{FID score (lower is better) is chosen as metric to compare the image fidelity of generated data for datasets DSB and BBBC039v1 with original images from the respective datasets}
\centering
    \begin{tabular}{lPP}
		\hline Model & \multicolumn{1}{c}{DSB} & \multicolumn{1}{c}{BBBC039v1} \\
		\hline Orig. CycleGAN & 110.1p3.0 & 380.2p12.9 \\
		w. DCT \bf{(ours)} & \bf{96.4}p\bf{0.7} & 328.8p13.8 \\
		w. noise \cite{noise} & 104.9p4.6 & 376.6p39.7 \\
		w. DnCNN \cite{porav_stego} & 101.3p3.5 & \bf{327.1}p\bf{4} \\
		\hline
	\end{tabular}
\label{tab:fidelity}
\end{table}
\subsection{Qualitative results of synthetic images}
\begin{figure}[htb]
  \centering
  \begin{subfigure}[b]{0.15\textwidth}
    \centering
    \includegraphics[width=2.5cm,height=2.5cm]{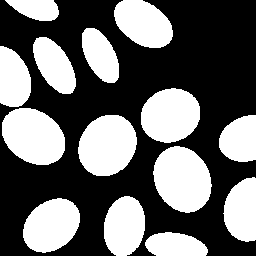}
  \end{subfigure}
  \hfill
  \begin{minipage}[b]{0.48\textwidth}
    \centering
    \subcaption{Original input mask}
    \label{fig:input_mask}
  \end{minipage}
  \vskip 0.1in
  \begin{subfigure}[b]{0.15\textwidth}
    \centering
    \caption*{\textbf{Generated Image}}
    \includegraphics[width=2.5cm,height=2.5cm]{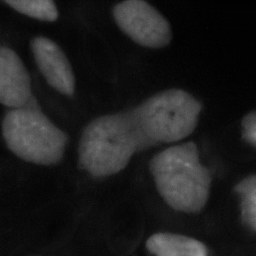}
  \end{subfigure}
  \begin{subfigure}[b]{0.15\textwidth}
    \centering
    \caption*{\textbf{Cycled Mask}}
    \includegraphics[width=2.5cm,height=2.5cm]{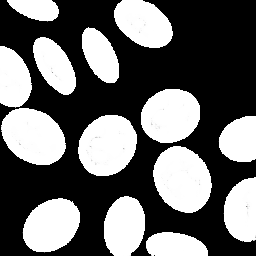}
  \end{subfigure}
  \begin{subfigure}[b]{0.15\textwidth}
    \centering
    \caption*{\textbf{Overlay}}
    \includegraphics[width=2.5cm,height=2.5cm]{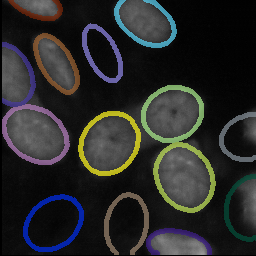}
  \end{subfigure}
  \hfill
  \begin{minipage}[b]{0.45\textwidth}
    \centering
    \subcaption{Original CycleGAN}
    \label{fig:second_row_caption}
  \end{minipage}
  \vskip 0.1in
  \begin{subfigure}[b]{0.15\textwidth}
    \centering
    \includegraphics[width=2.5cm,height=2.5cm]{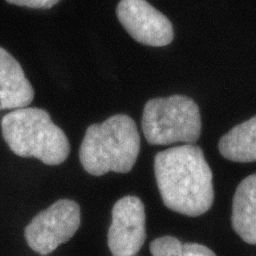}
  \end{subfigure}
  \begin{subfigure}[b]{0.15\textwidth}
    \centering
    \includegraphics[width=2.5cm,height=2.5cm]{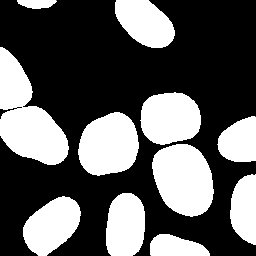}
  \end{subfigure}
  \begin{subfigure}[b]{0.15\textwidth}
    \centering
    \includegraphics[width=2.5cm,height=2.5cm]{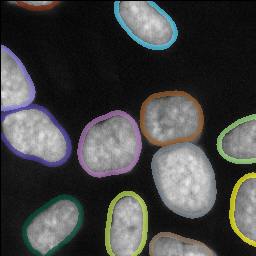}
  \end{subfigure}
  \hfill
  \begin{minipage}[b]{0.45\textwidth}
    \centering
    \subcaption{with DCT filtering \bf{(ours)}}
    \label{fig:third_row_caption}
  \end{minipage}
  \vskip 0.1in
  \begin{subfigure}[b]{0.15\textwidth}
    \centering
    \includegraphics[width=2.5cm,height=2.5cm]{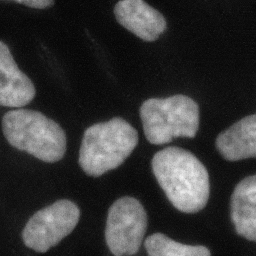}
  \end{subfigure}
  \begin{subfigure}[b]{0.15\textwidth}
    \centering
    \includegraphics[width=2.5cm,height=2.5cm]{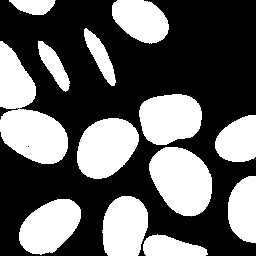}
  \end{subfigure}
  \begin{subfigure}[b]{0.15\textwidth}
    \centering
    \includegraphics[width=2.5cm,height=2.5cm]{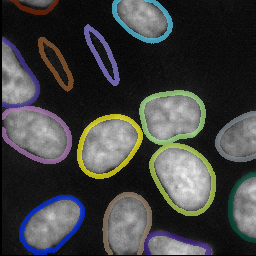}
  \end{subfigure}
  \hfill
  \begin{minipage}[b]{0.45\textwidth}
  \centering
    \subcaption{with \textit{DnCNN} \cite{porav_stego}}
    \label{fig:fourth_row_caption}
  \end{minipage}
  \vskip 0.1in
  \begin{subfigure}[b]{0.15\textwidth}
    \centering
    \includegraphics[width=2.5cm,height=2.5cm]{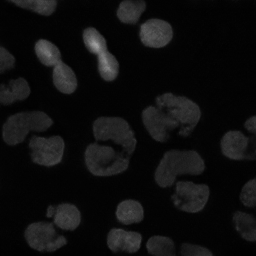}
  \end{subfigure}
  \begin{subfigure}[b]{0.15\textwidth}
    \centering
    \includegraphics[width=2.5cm,height=2.5cm]{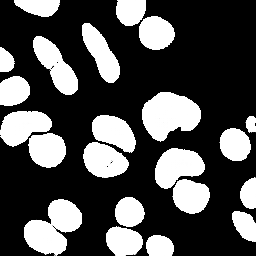}
  \end{subfigure}
  \begin{subfigure}[b]{0.15\textwidth}
    \centering
    \includegraphics[width=2.5cm,height=2.5cm]{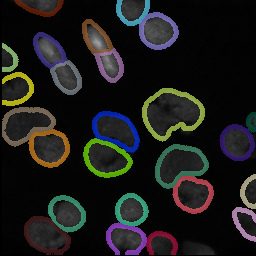}
  \end{subfigure}
  \hfill
  \begin{minipage}[b]{0.45\textwidth}
    \centering
    \subcaption{with Gaussian noise injection \cite{noise}}
    \label{fig:fifth_row_caption}
  \end{minipage}
  \vskip 0.1in
  \caption{Generated images based on the DSB dataset along with the cycled mask. The final segmentation mask used for StarDist training is shown in the third column as an overlay over the generated image.}
  \label{fig:results_comparison}
\end{figure}

A representative sample of generated images using the discussed methods is depicted along with (from left to right) the generating input mask, the respective cycled mask, and the post-processed instance mask as an overlay in Figure~\ref{fig:results_comparison}. For all four analyzed methods, we observe a deviation from the original input mask. This deviation is not necessarily negative as long as it is represented in the cycled mask. This is not the case for the unmodified CycleGAN. Multiple nuclei instances are not rendered into the image but occur in the cycled mask. n this case, it is almost a pixel-perfect copy of the original input mask. The reason for this is the before-mentioned CycleGAN steganography. If our proposed DCT filtering is used during training, we observe no steganography during test time. The cycled mask resembles the generated image closely. We also observe that the nuclei do not exhibit an exact ellipse shape, which we consider a positive effect. This allows the model to adapt to the shape of nuclei found in the training set and overcome shortcomings of the ellipse synthesis representations. For the approaches from literature, specifically the filtering with the denoising network \textit{DnCNN} \cite{porav_stego}, we observe that steganography was reduced but not fully removed as there are nuclei in the cycled mask that were not present in the generated image. For the noise injection approach from \cite{noise}, we observe the largest deviation from the input mask in the generated image. However, since this is reflected in the cycled mask, we conclude that removing the steganography was successful. We also observe a slightly less realistic reproduction of the typical nuclei texture found in the remaining images when noise injection is used.   
\subsection{Segmentation performance on DSB}
The results for the segmentation network tested on the real-data test split textit{stage1\_test} of the DSB dataset are reported in Table~\ref{tab:res_dsb} for an IoU threshold of $\tau = 0.5$ and in Table~\ref{tab:res_dsb_75} for an IoU threshold $\tau = 0.75$. The best results in terms of precision are achieved with the training set created by our proposed CycleGAN with DCT filtering, outperforming the baseline CycleGAN by $19.7$ percentage points for $\tau=0.5$ and $38.7$ percentage points for $\tau=0.75$. For the approach using \textit{DnCNN} we observe comparable performance at $\tau=0.5$, with a larger gap ($8.3$ percentage points less precision compared to DCT) for the more challenging threshold of $\tau=0.75$. The noise-injection approach suffers from a high variance leading to a slightly worse average performance.
\begin{table}[htbp]
\caption{StarDist segmentation performance in \% on DSB \textit{stage1\_test} split, trained on synthetic datasets generated from different CycleGAN methods. Scores are computed for $\tau = 0.5$}
\centering
\begin{tabular}{lPPP}
\hline
Model & \multicolumn{1}{c}{Precision} & \multicolumn{1}{c}{Recall} & \multicolumn{1}{c}{F1-Score} \\
\hline  Orig. CycleGAN & 74.3p5 & 63.7p3 & 68.6p4 \\
w. DCT \bf{(ours)} & 94.0p1 & 61.1p1 & 74.0p1 \\
w. noise \cite{noise} & 87.0p14 & 39.8p31 & 50.3p37 \\
w. DnCNN \cite{porav_stego} & 92.3p1 & 62.8p3 & \bf{74.7}p\bf{2}
\end{tabular}
\label{tab:res_dsb}
\end{table}
\begin{table}[htbp]
\caption{StarDist segmentation performance in \% on DSB \textit{stage1\_test}, trained on synthetic datasets generated from different CycleGAN methods. Scores are computed for $\tau = 0.75$, which requires more precise predictions compared to $\tau = 0.5$}
\centering
\begin{tabular}{lPPP}
\hline
Model & \multicolumn{1}{c}{Precision} & \multicolumn{1}{c}{Recall} & \multicolumn{1}{c}{F1-Score} \\
\hline Orig. CycleGAN & 49.9p5 & 40.2p3 & 43.3p4  \\
w. DCT (\bf{ours}) & 85.6p1 & 55.6p1 & \bf{67.4}p\bf{1}  \\
w. noise \cite{noise} & 67.7p32 & 35.5p29 & 44.5p36  \\
w. DnCNN \cite{porav_stego} & 77.3p11 & 52.4p5 & 62.4p7  \\
\end{tabular}
\label{tab:res_dsb_75}
\end{table}
\subsection{Segmentation performance on BBC039v1}
We report results for the downstream segmentation task on the BBBC039v1 test set for $\tau = 0.5$ and $\tau = 0.75$ in Table~\ref{tab:res_3bc} and Table~\ref{tab:res_3bc_75} respectively. We observe a high precision for all methods except the noise injection, indicating a low number of false positives and a high number of true positives. Compared to the DSB dataset, we observe higher scores among all methods and metrics due to the lower diversity and therefore complexity of the data set. The lower complexity results in a lower amount of steganography. Consequently, the methods differ only slightly in their F1-score, with DCT filtering being the best method by a margin of 2 percentage points ($\tau = 0.5$) compared to the baseline. The margin between the proposed \gls{DCT} filtering and the baseline CycleGAN methods increases for a higher threshold ($\tau = 0.75$) to $8.4$ percentage points in terms of F1-score. This indicates that DCT filtering leads to more realistic nuclei shapes. Similar to the results of the DSB dataset, we observe a high variability in performance for the noise-injection results. The reason of this is liekly due to noise injection causing the GAN training to be less stable, generating more deficient samples.
\begin{table}[htbp]
\caption{StarDist segmentation performance in \% on BBBC039v1 test set, trained on synthetic datasets generated from different CycleGAN methods. Scores are computed for $\tau = 0.5$}
\centering
\begin{tabular}{lPPP}
\hline
Model & \multicolumn{1}{c}{Precision} & \multicolumn{1}{c}{Recall} & \multicolumn{1}{c}{F1-Score} \\
\hline Orig. CycleGAN & 99.9p1 & 86.3p6 & 92.1p4 \\
w. DCT (\bf{ours}) & 97.9p1 & 90.7p1 & \bf{94.1}p\bf{1} \\
w. noise \cite{noise} & 82.1p29 & 76.3p28 & 79.1p28 \\
w. DnCNN \cite{porav_stego} & 99.9p1 & 82.4p2 & 90.3p2 \\
\end{tabular}
\label{tab:res_3bc}
\end{table}
\begin{table}[htbp]
\caption{StarDist segmentation performance on BBBC039v1 test set, trained on synthetic datasets generated from different CycleGAN methods. Scores are computed for $\tau = 0.75$}
\centering
\begin{tabular}{lPPP}
\hline
Model & \multicolumn{1}{c}{Precision} & \multicolumn{1}{c}{Recall} & \multicolumn{1}{c}{F1-Score} \\
\hline Orig. CycleGAN & 82.8p11 & 72.6p15 & 77.3p13  \\
w. DCT (\bf{ours}) & 89.2p10 & 82.5p8 & \bf{85.7}p\bf{9}  \\
w. noise \cite{noise} & 65.2p54 & 61.1p50 & 61.1p52  \\
w. DnCNN \cite{porav_stego} & 91.4p11 & 75.5p10 & 82.6p10  \\
\end{tabular}
\label{tab:res_3bc_75}
\end{table}
\section{Discussion}
In this work, we proposed a new anti-steganography approach for CycleGANs and evaluated existing methods to improve CycleGAN consistency for \gls{fm} nuclei generation. We further showed the performance of nuclei segmentation trained solely on synthetic data. Using our proposed method, we observe a qualitative improvement in the generated images and masks and outperform existing anti-steganography approaches quantitatively on a downstream segmentation task. A considerable advantage of DCT filtering compared to using a \textit{DnCNN} is the reduced computational overhead. Instead of feeding the generated images through a neural network (\ie, the \textit{DnCNN}), we make use of the efficient and computationally cheap implementation of the DCT algorithm. By pairing the cycled mask with the generated image for synthetic dataset creation, we allow the network to deviate from the input mask, allowing CycleGAN to overcome potential shortcomings of the mask synthesis model. This can be a crucial factor in enabling CycleGAN to learn the desired mapping. In \cite{boehland} B\"ohland \etal show that CycleGAN can only produce satisfactory results if the shape and occurrence of synthetic masks are similar to the ground-truth distribution. In unsupervised settings, the ground-truth distribution is not available, and estimating the parameters of the mask synthesis model relies on a limited number of samples, which can introduce challenges and uncertainties. A limitation of using the cycled masks is the necessity of introducing post-processing steps  to convert them into instance masks. This in turn might induce an unknown bias into the processed masks and is thus prone to errors. This could be avoided by directly generating instance masks in CycleGAN, \eg, by rendering additional boundaries of masks or using a StarDist-like mask formulation based on star-convex polygons. \\
In this work, we did not consider the segmentation capabilities of CycleGAN itself. Due to its architectural design, CycleGAN is not only trained to generate microscopy images from masks but also to segment microscopy images into masks. In future work, we want to explore to what extent this allows this allows CycleGAN to be used as an unsupervised nuclei segmentation network, \eg, by using segmentation losses such as the Dice loss for cycle consistency. Exploring the generalizability of the proposed steganography removal approach to other cycle-loss-based image-to-image translation tasks could open up new possibilities for improving various GAN-based applications.
\newpage
\section{Acknowledgments}
\balance
We gratefully acknowledge funding by the Deutsche Forschungsgemeinschaft (DFG, German Research Foundation) - Projektnummer 405969122 (J.\,U., S.\,U., K.\,B.) and Dhip campus - Bavarian aim (K.\,B., M.\,S., T.\,W.) M.\,T., F.\,W., M.\,G., and S.\,U. received funding from the European Research Council (ERC) under the European Union's Horizon 2020 research and innovation
program (ERC Grant No. 810316).

{\small
\bibliographystyle{ieee_fullname}
\bibliography{additional_lit}
}

\end{document}